\documentclass[10pt,conference]{IEEEtran}

\usepackage{cite}      

\usepackage{graphicx}  

\usepackage{psfrag}    

\usepackage{subfigure} 

\usepackage{url}       

\usepackage{amssymb}
\usepackage{amsmath}   
\usepackage{array}
\newtheorem{theorem}{Theorem}[section]

\newtheorem{proposition}{Proposition}[section]

\renewcommand{\QED}{\QEDopen}

\begin{document}
%
\title{n-Channel Asymmetric Multiple-Description\\ Lattice Vector Quantization}
%
%
\author{%
\authorblockN{Jan \O stergaard, Richard Heusdens, and Jesper Jensen}
\authorblockA{Delft University of Technology\\
Mekelweg 4, 2628CD, Delft,\\
The Netherlands\\
Email: \{j.ostergaard,r.heusdens,j.jensen\}@ewi.tudelft.nl}
}

%



\maketitle

\begin{abstract}
We present analytical expressions for optimal entropy-constrained multiple-description lattice vector quantizers which, under high-resolutions assumptions, minimize the expected distortion for given packet-loss probabilities. 
We consider the asymmetric case where packet-loss probabilities and side entropies are allowed to be unequal and find optimal quantizers for any number of descriptions in any dimension. 
We show that the normalized second moments of the side-quantizers are given by that of an $L$-dimensional sphere independent of the choice of lattices.
Furthermore, we show that the optimal bit-distribution among the descriptions is not unique. In fact, within certain limits, bits can be arbitrarily distributed.
\end{abstract}



%

\section{Introduction}
%
%
%
%
\vspace{-2mm}
Multiple-description coding (MDC) aims at creating separate descriptions individually capable of reproducing a source to a specified accuracy and when combined being able to refine each other. 
Traditionally quantizer based MDC schemes consider only two descriptions~\cite{vaishampayan:1993,vaishampayan:1994,vaishampayan:1998c,servetto:1999,diggavi:2000,vaishampayan:2001,diggavi:2002,goyal:2002,kelner:2000,ostergaard:2004}. Among the few vector quantizer based approaches which consider more than two descriptions are~\cite{fleming:1999,fleming:2004,ostergaard:2005c,ostergaard:2004b}. 
In~\cite{ostergaard:2005c,ostergaard:2004b} closed form expressions for the design of lattice vector quantizers are given for the symmetric case where all packet-loss probabilities and side entropies are equal. 
\vspace{-2mm}
\begin{figure}[ht]
\psfrag{X}{$\scriptstyle X$}
\psfrag{Xh}{$\scriptstyle \hat{X}$}
\psfrag{R0}{$\scriptstyle R_0$}
\psfrag{R1}{$\scriptstyle R_1$}
\psfrag{RK}{$\scriptstyle R_{K-1}$}
\psfrag{Description 0}{\scriptsize Description 0}
\psfrag{Description 1}{\scriptsize Description 1}
\psfrag{Description K-1}{\scriptsize Description $\scriptstyle K-1$}
\begin{center}
\includegraphics[width=9cm]{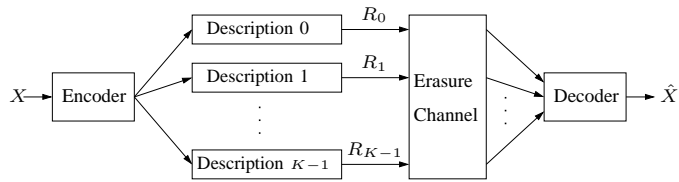}
\caption{General $K$-channel system. Descriptions are encoded at an entropy of $R_i$, $i=0,\dots, K-1$. The erasure channel either transmits the $i$th description errorless or not at all.}
\label{fig:nchannel}
\end{center}
\end{figure}
\vspace{-3mm}
In~\cite{fleming:1999,fleming:2004} iterative vector quantizer design algorithms are proposed for the asymmetric case where packet-loss probabilities and side entropies are allowed to be unequal. 

In this paper we consider the asymmetric case for an arbitrary number of descriptions, where the $i$th description is encoded at an entropy of $R_i$, for $i=0,\dots, K-1$, see Fig.~\ref{fig:nchannel}. The total rate is then given by the sum of the entropies of the individual descriptions.
Due to the asymmetry, the total distortion depends not only on how many descriptions are received (as is the case in the symmetric situation~\cite{ostergaard:2005c,ostergaard:2004b}), but also on \emph{which} descriptions make it to the decoder. 
We derive analytical expressions for the central and side quantizers which, under high-resolution assumptions, minimize the \emph{expected distortion} at the receiver subject to entropy constraints on the total rate.
In contrast to~\cite{fleming:1999,fleming:2004} our design allows for simple adaptation of our quantizers to changing source-channel characteristics and entropy constraints, effectively avoiding iterative quantizer design procedures.

\vspace{-1mm}
\section{Preliminaries}\label{sec:prelim}
\vspace{-1mm}
Let $X\in \mathbb{R}^L$ be an arbitrary i.i.d.\ source and let $\Lambda\subset \mathbb{R}^L$ be a real lattice with Voronoi regions $V(\lambda), \lambda\in \Lambda$, given by
\vspace{-1mm}
\begin{equation*}
V(\lambda) \triangleq \{ x\in \mathbb{R}^L : \| x - \lambda\|^2 \leq \|x-\lambda'\|^2,\, \forall\, \lambda' \in \Lambda \},
\end{equation*}
where $x$ is a realization of $X$ and we define $\|x\|^2 = \frac{1}{L}x^Tx$, where $T$ denotes vector transposition.

We consider one central lattice (central quantizer) $\Lambda_c$ and several sublattices (side quantizers) $\Lambda_i$, where $i=0,\dots, K-1$ and $K>0$, is the number of descriptions. The trivial case $K=1$ leads to a single-description system, where we would simply use one central quantizer and no side quantizers. 
We assume that sublattices are geometrically similar to $\Lambda_c$, i.e.\ they can be obtained from $\Lambda_c$ by applying change of scales, rotations and possible reflections. 
The sublattice index $N_i=[\Lambda_c:\Lambda_i], N_i\in \mathbb{Z}^+$, of the $i$th sublattice $\Lambda_i$ describes the volume $\nu_i$ of a sublattice cell relative to the volume $\nu$ of a central lattice cell. 
The volume $\nu_i$ of the $i$th sublattice cell is then given by $\nu_i=N_i\nu$. In the design of the index assignment map, we make use of a product lattice $\Lambda_\pi \subseteq \Lambda_i \subseteq \Lambda_c$, which is simply a sublattice of index $N_\pi = [\Lambda_c : \Lambda_\pi]$. 
To simplify the design of the index assignment map we assume sublattices are clean~\cite{conway:1999b}, specifically we require that no points of $\Lambda_c$ lies on the boundaries of the Voronoi regions of $\Lambda_\pi$.

\vspace{-2mm}
\subsection{Index assignments}\label{sec:index}
A source vector $x$ is quantized to the nearest reconstruction point $\lambda_c$ in the central lattice $\Lambda_c$. Hereafter follows index assignments (mappings), which uniquely map all $\lambda_c$'s to reconstruction points in each of the sublattices $\Lambda_i$. This mapping is done through a labeling function $\alpha$, and we denote the individual component functions of $\alpha$ by $\alpha_i$. In other words, the injective map $\alpha$ that maps $\Lambda_c$ into $\Lambda_0 \times \dots \times \Lambda_{K-1}$, is given by
\begin{equation*}
\alpha(\lambda_c)=(\alpha_0(\lambda_c),\alpha_1(\lambda_c),\dots,\alpha_{K-1}(\lambda_c)),
\end{equation*}
where $\alpha_i(\lambda_c)=\lambda_i \in \Lambda_i$ and $i=0,\dots, K-1$. Each $K$-tuple $(\lambda_0,\dots,\lambda_{K-1})$ is used only once when labeling points in $\Lambda_c$ in order to make sure that $\lambda_c$ can be recovered unambiguously when all $K$ descriptions are received. 

Since lattices are infinite arrays of points, we adopt the procedure used in~\cite{servetto:1999,diggavi:2000,vaishampayan:2001,diggavi:2002,ostergaard:2005c,ostergaard:2004b} and construct a shift invariant labeling function, so only a finite number of points must be labeled. We generalize the approach of~\cite{diggavi:2000,diggavi:2002} and construct a product lattice $\Lambda_\pi$ which has $N_\pi$ central lattice points and $N_\pi/N_i$ sublattice points from the $i$th sublattice in each of its Voronoi regions. The Voronoi regions $V_\pi$ of the product lattice $\Lambda_\pi$ are all similar so by labeling only central lattice points within one Voronoi region of $\Lambda_\pi$, the rest of the central lattice points may be labeled simply by translating this Voronoi region throughout $\mathbb{R}^L$. Without loss of generality, we let $N_\pi=\prod_{i=0}^{K-1} N_i$ and by construction we let $\Lambda_\pi$ be a geometrical similar and clean sublattice of $\Lambda_i$ as well as $\Lambda_c$.
With this choice of $\Lambda_\pi$, we only label central lattice points within $V_\pi(0)$, which is the Voronoi region of $\Lambda_\pi$ around origo. With this we get the following shift invariant property
\vspace{-2mm}
\begin{equation*}
\alpha(\lambda_c + \lambda_\pi) = \alpha(\lambda_c) + \lambda_\pi,
\end{equation*}
for all $\lambda_\pi \in \Lambda_\pi$ and all $\lambda_c \in \Lambda_c$. 

\subsection{Rate and distortion performance}
Using standard high-resolution assumptions for lattice quantizers~\cite{gray:1990}, the expected central distortion can be expressed as
\begin{equation}\label{eq:d0G}
d_c \approx G(\Lambda_c)\nu^{2/L},
\end{equation}
where $G(\Lambda_c)$ is the normalized second moment of inertia~\cite{conway:1999} of the central quantizer and it can be shown that the side distortion for the $i$th description is given by~\cite{ostergaard:2004b}
\vspace{-2mm}
\begin{equation}\label{eq:di}
d_i\approx d_c +  \frac{1}{N_\pi}\sum_{\lambda_c \in V_\pi(0)} \|\lambda_c - \alpha_i(\lambda_c)\|^2.
\end{equation}

The minimum entropy $R_c$ needed to achieve the central distortion $d_c$ is given by~\cite{gray:1990}
\vspace{-2mm}
\begin{equation}\label{eq:Rc}
R_c \approx h(X) - \frac{1}L\log_2(\nu),
\end{equation}
where $h(X)$ is the component-wise differential entropy of the source.
The side entropies are given by~\cite{ostergaard:2004b}
\vspace{-2mm}
\begin{equation}\label{eq:Ri}
R_i\approx h(X) - \frac{1}L\log_2(N_i\nu).
\end{equation}

\section{Construction of labeling function}
\label{sec:label}
The index assignment is done by a labeling function $\alpha$, that maps central lattice points to sublattice points. An optimal assignment minimizes the expected distortion when $1\leq \kappa \leq K-1$ descriptions are received and is invertible so the central quantizer can be used when all descriptions are received. 

\subsection{Expected distortion}
\label{sec:reconstruction}
At the receiving side, $X\in \mathbb{R}^L$ is reconstructed to a quality that is determined by the received descriptions. If no descriptions are received we reconstruct using the expected value, $E[X]$, and if all $K$ descriptions are received we reconstruct using the inverse map outlined above, hence obtaining the quality of the central quantizer. 
In all other cases, we reconstruct to the average\footnote{%
The average value of the received descriptions is equivalent to their centroid, since the pdf of $X$, under high-resolution assumptions, is constant within the region where elements of a $K$-tuple are located.}
of the received descriptions.

There are in general several ways of receiving $\kappa$ out of $K$ descriptions. Let $\mathcal{L}$ denote an index set consisting of all possible $\kappa$ combinations out of $\{0,\dots, K-1\}$ so that $|\mathcal{L}| = \binom{K}{\kappa}$. We denote an element of $\mathcal{L}$ by $l=\{l_0, \dots , l_{\kappa-1}\}$. The complement $l^c$ of $l$ denotes the $K-\kappa$ indices not in $l$, i.e.\ $l^c = \{0,\dots, K-1\}\backslash\{l\}$.
We will use the notation $\mathcal{L}_{i}$ to indicate the set of all $l\in \mathcal{L}$ that contains the index $i$, i.e., $\mathcal{L}_i=\{ l : l \in \mathcal{L}\ \text{and}\ i\in l\}$ and similarly $\mathcal{L}_{i,j} = \{ l : l \in \mathcal{L}\ \text{and}\ i,j\in l\}$.
Furthermore, let $p_i$ be the packet-loss probability for the $i$th description and, consequently, let $\mu_i=1-p_i$ be the probability that the $i$th description is received. Finally, let $p(l)=\prod_{i\in l}\mu_i \prod_{j\in l^c}p_j$, $p(\mathcal{L}) = \sum_{l\in\mathcal{L}} p(l)$, $p(\mathcal{L}_i) = \sum_{l\in\mathcal{L}_i} p(l)$ and $p(\mathcal{L}_{i,j}) = \sum_{l\in\mathcal{L}_{i,j}} p(l)$. For example for $K=3$ and $\kappa=2$ we have $\mathcal{L}=\{ \{0,1\}, \{0,2\}, \{1,2\}\}$ and hence $p(\mathcal{L})=\mu_0\mu_1 p_2 + \mu_0\mu_2 p_1 + \mu_1\mu_2 p_0$. 

Upon reception of any $\kappa$ out of $K$ descriptions we reconstruct $X$ as $\hat{X}=\frac{1}{\kappa}\sum_{j\in l}\lambda_j$ where the resulting distortion can be written similar to~(\ref{eq:di}), e.g.\ if descriptions $i$ and $j$ are received, the norm in~(\ref{eq:di}) should read $\|\lambda_c-0.5(\alpha_i(\lambda_c)+\alpha_j(\lambda_c))\|^2$. It follows that the expected distortion is given by
\vspace{-2mm}
\begin{align}\notag
&d_a^{(K,\kappa)} \approx \sum_{l\in \mathcal{L}} p(l)\left(d_c + \frac{1}{N_\pi}\sum_{\lambda_c\in V_{\pi}(0)}\left\| \lambda_c - \frac{1}{\kappa}\sum_{j=0}^{\kappa-1}\lambda_{l_j} \right\|^2\right) \\ \label{eq:expdist}
&= p(\mathcal{L})d_c + \frac{1}{N_\pi}\sum_{\lambda_c\in V_{\pi}(0)}\sum_{l\in \mathcal{L}}p(l)\left\| \lambda_c - \frac{1}{\kappa}\sum_{j=0}^{\kappa-1}\lambda_{l_j} \right\|^2,
\end{align}
where $\lambda_{l_j}=\alpha_{l_j}(\lambda_c)$ and the two special cases $\kappa\in \{0,K\}$ are given by $d_a^{(K,0)}\approx E[\|X\|^2]\prod_{i=0}^{K-1}p_i$ and $d_a^{(K,K)}\approx d_c\prod_{i=0}^{K-1}\mu_i$.

\subsection{Cost functional}
From (\ref{eq:expdist}) we see that the distortion $d_a^{(K,\kappa)}$ may be split into two terms, one describing the distortion occurring when the central quantizer is used on the source, and one that describes the distortion due to the index assignment. An optimal index assignment minimizes the second term in (\ref{eq:expdist}) for all possible combinations of descriptions. 
We can rewrite this term using the following theorem
\begin{theorem}\label{theo:sums}
For any $1\leq \kappa \leq K$ we have
\begin{equation*}
\begin{split}
&\sum_{\lambda_c}\sum_{l\in \mathcal{L}}p(l)\left\| \lambda_c - \frac{1}{\kappa}\sum_{j=0}^{\kappa-1}\lambda_{l_j} \right\|^2\\
&=\sum_{\lambda_c}\bigg(p(\mathcal{L})\bigg\| \lambda_c - \frac{1}{\kappa p(\mathcal{L})}\sum_{i=0}^{K-1}p(\mathcal{L}_i)\lambda_i \bigg\|^2 \\
&\quad+\frac{1}{\kappa^2}\sum_{i=0}^{K-2}\sum_{j=i+1}^{K-1}
\left(\frac{p(\mathcal{L}_i)p(\mathcal{L}_j)}{p(\mathcal{L})}-p(\mathcal{L}_{i,j})\right)\|\lambda_i - \lambda_j\|^2\bigg).
\end{split}
\end{equation*}
\end{theorem}
\begin{proof}
See~\cite{ostergaard:2005e}.
\end{proof}

The cost functional to be minimized can then be written as
\begin{equation}\label{eq:costfunctional2}
\begin{split}
&J^{(K,\kappa)}= \frac{1}{N_\pi}
\sum_{\lambda_c\in V_\pi(0)}\bigg(p(\mathcal{L})\left\| \lambda_c - \frac{1}{\kappa p(\mathcal{L})}\sum_{i=0}^{K-1}\lambda_i p(\mathcal{L}_i)\right\|^2 \\
&+
\frac{1}{\kappa^2}\sum_{i=0}^{K-2}\sum_{j=i+1}^{K-1}
\|\lambda_i - \lambda_j\|^2
\left(\frac{p(\mathcal{L}_i)p(\mathcal{L}_j)}{p(\mathcal{L})}-p(\mathcal{L}_{i,j})\right)\bigg).
\end{split}
\end{equation}
We minimize this cost functional subject to a constraint on the sum of the side entropies. We remark here that the side entropies depend solely on $\nu$ and $N_i$ and as such not on the particular choice of $K$-tuples. In other words, for fixed $N_i$'s and a fixed $\nu$, the index assignment problem is solved if~(\ref{eq:costfunctional2}) is minimized. The problem of choosing $\nu$ and $N_i$ such that certain entropy constraints are not violated is independent of the assignment problem and deferred to Section~\ref{sec:optq}.

The first term in~(\ref{eq:costfunctional2}) describes the distance from a central lattice point to the weighted centroid of its associated $K$-tuple. The second term describes the weighted sum of pairwise squared distances (WSPSD) between elements of the $K$-tuples. It can be shown, c.f.\ Proposition~\ref{prop:growthriemann2}, that, under a high-resolution assumption, the second term in~(\ref{eq:costfunctional2}) is dominant, from which we conclude that 
in order to minimize~(\ref{eq:costfunctional2}) we must use $K$-tuples with the smallest WSPSD\@. 
These $K$-tuples are then assigned to central lattice points in such a way, that the first term in~(\ref{eq:costfunctional2}) is minimized. This problem can be posed and solved as a linear assignment problem~\cite{west:2001}.

\subsection{Minimizing cost functional}
\label{sec:ktuples}
To obtain $K$-tuples we center a region $\tilde{V}$ around all sublattice points $\lambda_0 \in \Lambda_0 \cap V_\pi(0)$, and construct $K$-tuples by combining sublattice points from the other sublattices (i.e.\ $\Lambda_i, i=1,\dots,K-1$) within $\tilde{V}(\lambda_0)$ in all possible ways and select the ones that minimize~(\ref{eq:costfunctional2}). For each $\lambda_0\in \Lambda_0\cap V_\pi(0)$ it is possible to construct $\prod_{i=1}^{K-1}\tilde{N}_i$ different $K$-tuples, where $\tilde{N}_i$ is the number of sublattice points from the $i$th sublattice within the region $\tilde{V}$. 
This gives a total of $(N_\pi/N_0)\prod_{i=1}^{K-1}\tilde{N}_i$ $K$-tuples when all $\lambda_0\in \Lambda_0 \cap V_\pi(0)$ are used. Let $\tilde{\nu}$ be the volume of $\tilde{V}$. Since $\tilde{N}_i=\tilde{\nu}/\nu N_i$ and we need $N_0$ $K$-tuples for each $\lambda_0\in V_\pi(0)$, we see that
\begin{equation*}
N_0 = \prod_{i=1}^{K-1}\tilde{N}_i = \frac{\tilde{\nu}^{K-1}}{\nu^{K-1}}\prod_{i=1}^{K-1}N_i^{-1},
\end{equation*}
so in order to obtain at least $N_0$ $K$-tuples, the volume of $\tilde{V}$ must satisfy
\begin{equation}\label{eq:vtilde}
\tilde{\nu}\geq \nu\prod_{i=0}^{K-1}N_i^{1/(K-1)}.
\end{equation}
For the symmetric case, i.e.\ $N=N_i$, $i=0,\dots,K-1$, we have $\tilde{\nu}\geq \nu N^{K/(K-1)}$, which is in agreement with the results obtained in~\cite{ostergaard:2005c,ostergaard:2004b}.

By centering $\tilde{V}$ around each $\lambda_0\in\Lambda_0\cap V_\pi(0)$, we make sure that the map $\alpha$ is shift-invariant. However, this also means that all $K$-tuples have their first coordinate (i.e.\ $\lambda_0$) inside $V_\pi(0)$. To be optimal this restriction must be removed which is easily done by considering all cosets of each $K$-tuple. The coset of a fixed $K$-tuple, say $t=(\lambda_0,\lambda_1,\dots,\lambda_{K-1})$ where $\lambda_0\in \Lambda_0\cap V_\pi(0)$, is given by $\mathrm{Coset}(t)=\{t+\lambda_\pi\}$, for all $\lambda_\pi \in \Lambda_\pi$. The $K$-tuples in a coset are distinct modulo $\Lambda_\pi$ and by making sure that only one member from each coset is used, the shift-invariance property is preserved. 

Before we outline the design procedure for constructing an optimal index assignment we remark that in order to minimize the WSPSD between a fixed $\lambda_i$ and the set of points $\{\lambda_j \in \Lambda_j\cap \tilde{V}\}$ it is required that $\tilde{V}$ forms a sphere centered at $\lambda_i$.
\begin{enumerate}
\item Center a sphere $\tilde{V}$ at each $\lambda_0\in \Lambda_0\cap V_\pi(0)$ and construct all possible $K$-tuples $(\lambda_0,\lambda_1,\dots,\lambda_{K-1})$ where $\lambda_i\in \Lambda_i\cap \tilde{V}(\lambda_0)$ and $i=1,\dots, K-1$. Notice that all $K$-tuples have their first coordinate ($\lambda_0$) inside $V_\pi(0)$ and they are therefore shift-invariant. Make $\tilde{V}$ large enough so at least $N_0$ distinct $K$-tuples are found for each $\lambda_0$.
\item Construct cosets of each $K$-tuple. 
\item The $N_\pi$ central lattice points in $\Lambda_c\cap V_\pi(0)$ must now be matched to distinct $K$-tuples. This is a standard linear assignment problem~\cite{west:2001} where only one member from each coset is (allowed to be) matched to a central lattice point in $V_\pi(0)$.
\end{enumerate}

As observed in~\cite{ostergaard:2004b}, having equality in~(\ref{eq:vtilde}), i.e.\ using the minimum $\tilde{\nu}$, will not minimize the WSPSD\@. Instead a slightly larger region must be used. 
For the \textit{practical} construction of the $K$-tuples this is not a problem, since we simply use e.g.\ twice as large a region as needed and let the linear assignment algorithm choose the optimal $K$-tuples. 
However, in order to \emph{theoretically} describe the performance of the quantizers we need to know the optimal $\tilde{\nu}$. In~\cite{ostergaard:2005c,ostergaard:2004b} an expansion factor $\psi$ was introduced and used to describe how much $\tilde{V}$ had to be expanded from the theoretical lower bound~(\ref{eq:vtilde}), to make sure that the $N_0$ optimal $K$-tuples could be constructed by combining sublattice points within the region $\tilde{V}$. 
Adopting this approach leads to $\tilde{\nu}=\psi\nu\prod_{i=0}^{K-1}N_i^{1/(K-1)}$ where e.g.\ for the two-dimensional case $\psi~\approx 2^{(K-2)/(K-1)}$\cite{ostergaard:2005c,ostergaard:2004b}. Analytical expressions for $\psi$ are given in~\cite{ostergaard:2005b}.

\section{High-resolution analysis}\label{sec:highrate}
In this section we derive high-resolution approximations for the expected distortion. However, we first introduce Proposition~\ref{prop:riemann2} which relates the sum of distances between pairs of sublattice points to $G(S_L)$, the dimensionless normalized second-moment of an $L$-dimensional sphere. Hereafter follows Proposition~\ref{prop:growthriemann2} which determines the dominating term in the expression for the expected distortion.

\begin{proposition}\label{prop:riemann2}
For $N_i\rightarrow \infty$ and $\nu_i\rightarrow 0$, we have
for any pair of sublattices, $(\Lambda_i,\Lambda_j),\ i,j=0,\dots,K-1,\ i\neq j$,
\begin{equation*}
\begin{split}
\sum_{\lambda_c\in V_\pi(0)}
\| \alpha_i(\lambda_c)&-\alpha_j(\lambda_c)\|^2 \\[-4mm]
&\approx \psi^{2/L}\nu^{2/L} G(S_L) N_\pi\prod_{m=0}^{K-1}N_m^{2/L(K-1)}.
\end{split}
\end{equation*}
\end{proposition}

\begin{proof}
Let $T_i = \{ \lambda_i : \lambda_i = \alpha_i(\lambda_c),\ \lambda_c \in V_\pi(0) \}$, i.e.\
the set of $N_\pi$ sublattice points $\lambda_i \in \Lambda_i$ associated with the $N_\pi$ central lattice points within $V_\pi(0)$.
Furthermore let $T'_i \subset T_i$ be the set of unique elements of $T_i$, where $|T'_i|\approx N_\pi/N_i$. 
Finally, let $T_j(\lambda_i) = \{ \lambda_j : \lambda_j = \alpha_j(\lambda_c)\ \text{and}\ \lambda_i = \alpha_i(\lambda_c),\  \lambda_c \in V_\pi(0) \}$ so 
that $T_j(\lambda_i)$ contains all the elements $\lambda_j\in \Lambda_j$ which are in the $K$-tuples that also contains a specific $\lambda_i \in \Lambda_i$. Let $T'_j(\lambda_j)\subset T_j(\lambda_i)$ be the set of unique elements. 

For sublattice $\Lambda_i$ and $\Lambda_j$ we have
\begin{equation*}
\sum_{\lambda_c\in V_\pi(0)}\| \alpha_i(\lambda_c) - \alpha_j(\lambda_c)\|^2 =
\sum_{\lambda_i\in T'_i}\sum_{\lambda_j\in T_j(\lambda_i)} \|\lambda_i - \lambda_j \|^2.
\end{equation*}
Observe that each $\lambda_i\in V_\pi(0)$ is used $N_\pi/|T'_i|\approx N_i$ times, so given $\lambda_i\in T'_i$, we have
\begin{equation*}
\begin{split}
\sum_{\lambda_j\in T_j(\lambda_i)} \|\lambda_i - \lambda_j \|^2 \nu_j
&\approx
\frac{N_i}{\tilde{N}_j}\sum_{\lambda_j\in T'_j(\lambda_i)}\|\lambda_i-\lambda_j\|^2\nu_j \\
&\approx \frac{N_i}{\tilde{N}_j} \int_{\tilde{V}(\lambda_i)}\|\lambda_i-x\|^2\, dx \\
&\approx \frac{N_i}{\tilde{N}_j} \tilde{\nu}^{1+2/L} G(S_L) \\
&= N_i \nu_j \tilde{\nu}^{2/L} G(S_L) \\
\end{split}
\end{equation*}
since $\tilde{N}_j=\tilde{\nu}/\nu_j$. Hence, with $\tilde{\nu}=\psi\nu\prod_{m=0}^{K-1}N_m^{1/(K-1)}$, we have
\begin{equation*}
\begin{split}
\sum_{\lambda_j\in T_j(\lambda_i)} \|\lambda_i &- \lambda_j \|^2 \nu_j  \\[-4mm]
&\approx
N_i\nu_j \psi^{2/L} \nu^{2/L}G(S_L)\prod_{m=0}^{K-1}N_m^{2/L(K-1)},
\end{split}
\end{equation*}
which is independent of $\lambda_i$, so that
\vspace{-2mm}
\begin{equation*}
\begin{split}
\sum_{\lambda_i\in T'_i}\sum_{\lambda_j\in T_j(\lambda_i)}\| \lambda_i &- \lambda_j\|^2 
\approx \frac{N_\pi}{N_i}\sum_{\lambda_j\in T_j(\lambda_i)}\| \lambda_i - \lambda_j\|^2 \\
&\approx \psi^{2/L} \nu^{2/L} G(S_L) N_\pi\prod_{m=0}^{K-1}N_m^{2/L(K-1)},
\end{split}
\end{equation*}
which completes the proof.
\end{proof}

\begin{proposition}\label{prop:growthriemann2}
Let $k$ be chosen such that $N_k\geq N_i$ for all $i,k \in\{0,\dots, K-1\}$. For $N_i\rightarrow \infty$ and $\prod_{\substack{m=0\\m\neq k}}^{K-1} N_m>N_k^{K-2}$ we have
\vspace{-1mm}
\begin{equation*}
\mathcal{O}\left(
\frac{\sum\limits_{\lambda_c}\left\| \lambda_c - \frac{1}{\kappa p(\mathcal{L})}\sum_{i=0}^{K-1}p(\mathcal{L}_i)\lambda_i \right\|^2}
{\sum\limits_{\lambda_c}\sum\limits_{i=0}^{K-2}\sum\limits_{j=i+1}^{K-1}
\left(\frac{p(\mathcal{L}_i)p(\mathcal{L}_j)}{p(\mathcal{L})}-p(\mathcal{L}_{i,j})\right)\| \lambda_i - \lambda_j\|^2 }\right) \rightarrow 0.
\end{equation*}
\end{proposition}
\begin{proof}
See~\cite{ostergaard:2005e}.
\end{proof}

The expected distortion (\ref{eq:expdist}) can by use of Theorem~\ref{theo:sums} be written as
\vspace{-2mm}
\begin{equation}\label{eq:da}
\begin{split}
&d_a^{(K,\kappa)}\approx p(\mathcal{L})\,d_c + \frac{1}{N_\pi}\sum_{\lambda_c\in V_{\pi}(0)}\sum_{l\in \mathcal{L}}p(l)\left\| \lambda_c - \frac{1}{\kappa}\sum_{j=0}^{\kappa-1}\lambda_{l_j} \right\|^2 \\
&=
p(\mathcal{L})\,d_c + 
\frac{1}{N_\pi}\sum_{\lambda_c\in V_\pi(0)} \bigg(p(\mathcal{L})\left\| \lambda_c - \frac{1}{\kappa p(\mathcal{L})}\sum_{i=0}^{K-1}p(\mathcal{L}_i)\lambda_i \right\|^2 \\
&+ \frac{1}{\kappa^2}\sum_{i=0}^{K-2}\sum_{j=i+1}^{K-1}
\left(\frac{p(\mathcal{L}_i)p(\mathcal{L}_j)}{p(\mathcal{L})}-p(\mathcal{L}_{i,j})\right)\|\lambda_i - \lambda_j\|^2\bigg).
\end{split}
\end{equation}

By use of Propositions~\ref{prop:riemann2},~\ref{prop:growthriemann2} and Eq.~(\ref{eq:d0G}) it follows that~(\ref{eq:da}) can be written as
\begin{equation*}
\begin{split}
d_a^{(K,\kappa)}&\approx G(\Lambda_c)\nu^{2/L}p(\mathcal{L})\\
&\quad +\psi^{2/L}\nu^{2/L} G(S_L)\beta\prod_{m=0}^{K-1}N_m^{2/L(K-1)},
\end{split}
\end{equation*}
where $\beta$ depends on $K$ and $\kappa$ and is given by
\begin{equation*}
\beta=\frac{1}{\kappa^2}
\sum_{i=0}^{K-2}\sum_{j=i+1}^{K-1}
\left(\frac{p(\mathcal{L}_i)p(\mathcal{L}_j)}{p(\mathcal{L})}-p(\mathcal{L}_{i,j})\right).
\end{equation*}
The total expected distortion is obtained by summing over $\kappa$ including the cases where $\kappa=0$ and $\kappa=K$,
\begin{equation}\label{eq:adistopt}
\begin{split}
d_a &\approx G(\Lambda_c)\nu^{2/L}\hat{p}(\mathcal{L}) + E[\|X\|^2]\prod_{i=0}^{K-1}p_i \\
&\quad +\psi^{2/L}\nu^{2/L} G(S_L)\prod_{m=0}^{K-1}N_m^{2/L(K-1)}\hat{\beta},
\end{split}
\end{equation}
where $\hat{p}(\mathcal{L})=\sum_{\kappa=1}^{K}p(\mathcal{L})$
and $\hat{\beta} = \sum_{\kappa=1}^{K}\beta$.

Using~(\ref{eq:Rc}) and~(\ref{eq:Ri}) we can write the expected distortion as a function of entropies, which leads to
\begin{equation}\label{eq:daopt}
\begin{split}
&d_a \approx G(\Lambda_c)2^{2(h(X)-R_c)}\hat{p}(\mathcal{L}) + E[\|X\|^2]\prod_{i=0}^{K-1}p_i\\
&+\psi^{2/L} \hat{\beta} G(S_L)2^{2(h(X)-R_c)}2^{\frac{2K}{K-1}\left(R_c - \frac{1}{K}\sum_{i=0}^{K-1}R_i\right)}.
\end{split}
\end{equation}

\section{Optimal quantizers}\label{sec:optq}
In this section we consider the situation where the total bit budget is constrained, i.e.\ we find the optimal scaling factors, $N_i$ and $\nu$, subject to entropy constraints on the sum of the side entropies $\sum_i R_i \leq R^*$, where $R^*$ is the target entropy. We also find the optimal bit-distribution among the $K$ descriptions. 

First we observe from~(\ref{eq:daopt}) that the expected distortion depends upon the \emph{sum} of the side entropies and not the individual side entropies. In order to be optimal it is necessary to achieve equality in the entropy constraint, i.e.\ $R^*=\sum_i R_i$. From~(\ref{eq:Ri}) we have
\begin{equation*}
\sum_{i=0}^{K-1}R_i = \sum_{i=0}^{K-1}h(X)-\frac{1}{L}\log_2(N_i\nu) = R^*,
\end{equation*}
which can be rewritten as
\begin{equation}\label{eq:tmptau}
\prod_{i=0}^{K-1}(N_i\nu) = 2^{L(Kh(X)-R^*)} = \tau_*,
\end{equation}
where $\tau_*$ is constant for fixed target and differential entropies. Writing~(\ref{eq:tmptau}) as
\begin{equation*}
\prod_{i=0}^{K-1}N_i^{2/L(K-1)}=\nu^{-2K/L(K-1)}\tau_*^{2/L(K-1)},
\end{equation*}
and inserting in~(\ref{eq:adistopt}) leads to
\begin{equation}\label{eq:edisttmp}
\begin{split}
d_a &\approx G(\Lambda_c)\nu^{2/L}\hat{p}(\mathcal{L}) + E[\|X\|^2]\prod_{i=0}^{K-1}p_i \\
&\quad +\psi^{2/L}\nu^{-2/L(K-1)}\tau_*^{2/L(K-1)} G(S_L)\hat{\beta}.
\end{split}
\end{equation}
The optimal $\nu$ is found by differentiating~(\ref{eq:edisttmp}) w.r.t.\ $\nu$, equating to zero and solving for $\nu$, which leads to
\vspace{-1mm}
\begin{equation}\label{eq:optnuRt}
\nu = 2^{L(h(X)-\frac{1}{K}R^*)}\left(\psi^{2/L}\frac{1}{K-1}\frac{G(S_L)}{G(\Lambda_c)}\frac{\hat{\beta}}
{\hat{p}(\mathcal{L})}\right)^{\frac{L(K-1)}{2K}}.
\end{equation}

At this point we still need to find expressions for the optimal $R_i$ (or, equivalently, optimal $N_i$ given $\nu$). Let $R_i = a_iR^*$, where $\sum_i a_i=1, a_i\geq 0$, hence $R^*=\sum_i R_i$. From~(\ref{eq:Ri}) we have
\vspace{-1mm}
\begin{equation*}
R_i=h(X)-\frac{1}L\log_2(N_i\nu)=a_iR^*,
\end{equation*}
which can be rewritten as
\vspace{-1mm}
\begin{equation*}
N_i=\nu^{-1}2^{L(h(X)-a_iR^*)},
\end{equation*}
where, after inserting the optimal $\nu$ from~(\ref{eq:optnuRt}) we obtain an expression for the optimal index value $N_i$, that is
\begin{equation*}
N_i = 2^{\frac{L}{K}(1-a_i)R^*}
\left(\psi^{-2/L}(K-1)\frac{G(\Lambda_c)}{G(S_L)}\frac{\hat{p}(\mathcal{L})}{\hat{\beta}}\right)^{\frac{L(K-1)}{2K}}.
\end{equation*}

It follows from~(\ref{eq:Ri}) that $R_c\geq a_iR^*$ so that $a_i\leq R_c/R^*$. In addition, since rates must be positive, we obtain the following inequalities
\vspace{-2mm}
\begin{equation}\label{eq:ai}
0<a_iR^* \leq R_c,\quad i=0,\dots,K-1.
\end{equation}
Hence, the individual side entropies $R_i=a_iR^*$ can be arbitrarily chosen as long as they satisfy~(\ref{eq:ai}) and $\sum_i a_i=1$. 

\vspace{-2mm}
\section{Results}
\begin{figure}[ht]
\begin{center}
\psfrag{p3}{$p_3$}
\includegraphics[width=8cm]{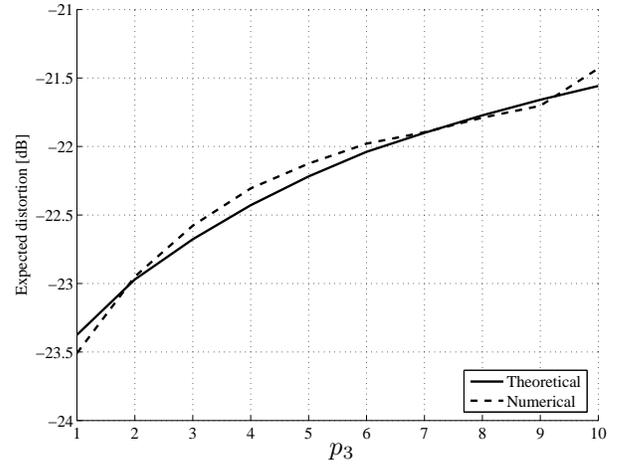}
\label{fig:a2K4perf}
\vspace{-2mm}
\caption{Expected distortion as a function of packet-loss probabilities.}
\label{fig:a2K4_perf}
\vspace{-6mm}
\end{center}
\end{figure}

\vspace{-1mm}
To verify theoretical results we present in this section experimental results obtained by using $2\cdot 10^6$ two-dimensional zero-mean unit-variance Gaussian source vectors.
Fig.~\ref{fig:a2K4_perf} shows the theoretical expected distortion~(\ref{eq:adistopt}) and the numerical expected distortion obtained for $K=4$ descriptions when using the $Z^2$ quantizer~\cite{conway:1999} at a total entropy $R^*=8$ bits/dimension.
In this setup we have $\psi=2^{(K-2)/(K-1)}$ and packet-loss probabilities are fixed at $p_0=2.5\%, p_1=5\%, p_2=7.5\%$ except for $p_3$ which is varied in the range $[1;10]\%$. As $p_3$ is varied we update $\nu$ according to~(\ref{eq:optnuRt}) and arbitrarily pick the index values $N_i$ such that $\sum_i R_i\leq R^*$. However, index values are restricted to a certain set of integers~\cite{diggavi:2002,ostergaard:2004b} and the side entropies might therefore not sum exactly to $R^*$. 
To make sure the target entropy is met with equality we then rescale $\nu$ as $\nu=2^{L(h(X)-\frac{1}K R^*)}\prod_{i=0}^{K-1}N_i^{-1/K}$. 
We see from Fig.~\ref{fig:a2K4_perf} a good correspondence between the theoretically and numerically obtained results.

\vspace{-2mm}
\section*{Acknowledgment}
\vspace{-1mm}
This research is supported by the Technology Foundation STW, applied science division of NWO and the technology programme of the ministry of Economics Affairs.




\bibliographystyle{IEEEtran.bst}

\end{document}